\documentclass{llncs}
\usepackage{graphicx}
\usepackage{amssymb}

\newcommand{\qin}{\hspace*{0.15in}}
\newenvironment{SProg}
     {\begin{small}\begin{tt}\begin{tabular}[t]{l}}%
     {\end{tabular}\end{tt}\end{small}}
\def\anno#1{{\ooalign{\hfil\raise.07ex\hbox{\small{\rm #1}}\hfil%
        \crcr\mathhexbox20D}}}

\title{A Web-based Tool Combining Different Type Analyses}
\author{Kim S. Henriksen
\and John P. Gallagher
}
\institute{Computer Science, Building 42.1, P.O. Box 260,\\
Roskilde University, DK-4000 Denmark\\
Email: {\tt \{kimsh,jpg\}@ruc.dk}
}
\begin{document}
\maketitle
\begin{abstract}
There are various kinds of type analysis of logic programs.
These include for example inference of types that describe an over-approximation
of the success set of a program,  
inference of well-typings, and 
abstractions based on given types.  Analyses can be descriptive or prescriptive or
a mixture of both,
and they can be goal-dependent or goal-independent.
We describe a prototype tool that can be accessed from a web browser, allowing various
type analyses to be run.  The first goal of the tool is to allow
the analysis results to be examined conveniently by clicking on points in the original
program clauses, and to highlight ill-typed program constructs, empty types or other type anomalies.
Secondly the tool allows combination of the various styles of analysis.  For example, a
descriptive regular type
can be automatically inferred for a given program, and then that type can be used 
to generate the minimal
``domain model" of the program with respect to the corresponding pre-interpretation, which can
give
more precise information than the original descriptive type.
\end{abstract}

\section{Introduction}

Type analysis has a long history in logic programming, going back to  \cite{Bruynooghe82a,Horiuchi-Kanamori,Mishra,Yardeni-Shapiro}.
Roughly speaking we can divide the field into descriptive typing and prescriptive
typing.  In the former, we start with an untyped program and infer a set of type rules
describing the terms that can appear in each argument position, automatically inventing types in
the process.   In the latter, we start from a program and 
some given types as well as possibly type signatures for the predicates, and infer types for the variables or other program
points in the
program. 

Using a simple example, let us review the different approaches to type analysis.
The descriptive approaches can be split into those that infer over-approximations, and those
that compute a well-typing. The difference can be seen in a simple program such as the {\em append}
program.  Using a descriptive over-approximation, we could obtain a type signature {\tt append(list,any,any)}
where the type {\tt list} is automatically constructed and defined by the regular type rules {\tt list --> []; [any|list]} where {\tt any} is the type containing
all terms.
Note that the second and third arguments cannot be approximated more precisely than {\tt any}
because (a) the base case of the {\em append} program allows any terms in these arguments, so that for example
{\tt append([],foo,foo)} is a consequence of the program, and (b) because the abstraction is based on arguments
rather than relations.  This means that we cannot capture the fact that the second argument of the program
is a list if and only if the third argument is also a list.  

A well-typing \cite{BruynoogheGH05} could infer a type
signature  {\tt append(list(X),list(X),list(X))}, where {\tt list(X)} is defined by the parametrised
type rule
 {\tt list(X) --> []; [X|list(X)]}, or else simply  {\tt append(list,list,list)} where {\tt list} is
defined as above.  This is not an over-approximation
of the success-set of the program; for example {\tt append([],foo,foo)} is not well-typed by this
signature.  However, it gives a well-typing to all uses of the {\em append} program that are in fact
used to append lists.

Now consider an approach in which types are given.  Given the types {\tt list} and {\tt any} as defined above,
an abstraction can be automatically constructed in which each term is interpreted either as {\tt list} or as
{\tt nonlist}, where {\tt nonlist} is defined as the difference of {\tt any} and {\tt list}. (This is equivalent
to a pre-interpretation over the domain {\tt \{list, nonlist\}} \cite{BoBD94,BoB94,Gallagher-Boulanger-Saglam-ILPS95}.) Computing the
least model with respect to this abstraction, we obtain a model {\tt \{append(list,list,list), append(list,nonlist,nonlist)}\}.
This is a relational over-approximation of the success set of the {\em append} program. It is more accurate than
{\tt append(list,any,any)}; for example the atom {\tt append([],foo,[])} is allowed by {\tt append(list,any,any)} 
but not by the model {\tt \{append(list,list,list), append(list,nonlist,nonlist)}\}.

The combination of these styles has not been investigated much to our knowledge, but the
possibilities seem interesting.  For instance, well-typings can be found in time roughly linear in the program
being analysed \cite{BruynoogheGH05}.  However, these are not safe approximations and cannot directly be used to
find type errors.  However, by using the inferred types prescriptively to compute an abstract model, we can
find bugs such as predicates that have no solution in the model, or variables that cannot be assigned a non-empty
type.  Similarly, an analysis based on a domain of NFTAs (non-deterministic finite tree automata)  can yield complex 
types that would be
very hard for a programmer to write by hand.  These types, when used in a prescriptive analysis, can  
capture
deeper properties of the program under analysis than could be easily identified otherwise.

In Section \ref{methods} we summarise the different type analysis methods handled by the tool.
Section \ref{qa} briefly describes the approach to goal-dependent analyses in a bottom-up,
goal-independent
analysis framework. Section \ref{sec:desctopresctypes} presents the approach to using descriptive types
as input to prescriptive type analyses, in order to gain precision.  Section \ref{impl} outlines the 
implementation of the tool and its web interface.  Section \ref{future} discusses some future work
and gives a summary and conclusion.

\section{Type Analysis Methods}\label{methods}
The type analysis tool includes at present three main type analysis engines:  
\begin{itemize}
\item
Domain model: given a program and a regular type, the tool computes the least model with respect to the 
pre-interpretation derived from the type. A BDD-based solver can be optionally used to compute the model,
giving greater scalability (at the cost of some pre-processing).
\item
Well-typing:  given a program, the tool computes types and predicate signatures that are a well-typing for the program.
\item
NFTA:  given a program, the tool computes a regular type (non-deterministic finite tree automaton) that
represents an over-approximation of the least model of the program.
\end{itemize}

This is not by any means a complete set of type analysis tools. For instance, we did not include a type analysis
tool using widening \cite{Gallagher-deWaal-ICLP94,VanHentenryck-Cortesi-LeCharlier,VaucheretB02}, or a polymorphic type analysis with explicit type union and intersection
operations \cite{Lu00}.  These could be incorporated later, but we believe that the above tools are
representative of the main styles of type analysis. These tools are all goal-independent but  we have implemented
goal-dependent analyses using query-answer transformations \cite{Debray-Ramakrishnan-94,Gallagher-deWaal-LOPSTR92}.
We review next the main aspects of each of these tools.

\subsection{Domain Model}\label{dm}

The analysis is based on given regular types.  More details can be
found in \cite{Gallagher-Henriksen-ICLP04,Gallagher-Henriksen-Banda-2005}. A
regular type $R$ can be viewed as a finite tree automaton (FTA) \cite{Comon}, in which a type
rule say {\tt list --> []; [dynamic|list]} is viewed as the tree automata transitions $[] \rightarrow {\it list}$
and $[{\it dynamic}|{\it list}] \rightarrow {\it list}$\footnote{The name 
{\tt dynamic} is used for the type of all terms and is a synonym for {\tt any} .  The name is chosen due to the connection with binding time analysis in partial
evaluation \cite{Jones-Gomard-Sestoft,CraigGLH04}.  It is defined by the set of type rules of the form {\tt f(dynamic,...,dynamic) -> dynamic} for each function
{\tt f} in the signature of the language.}.  Using a standard algorithm from
tree automaton theory we can obtain an FTA $R'$ that accepts exactly the same set of terms 
as $R$, but is {\em bottom-up deterministic} (a DFTA).  Each rule of a DFTA is of the form $f(q_1,\ldots,q_n) \rightarrow q$
and there are no two terms with the same left hand side. This implies that each term is accepted by
at most one state of $R'$.  Determinisation of an FTA with states $Q$ yields an FTA whose states are a subset of $2^Q$.
A state $\{q_1,\ldots,q_k\}$ in the DFTA accepts terms that are accepted by all of
the states $q_1,\ldots,q_k$ in the original FTA, and by no other states in $Q$.
For instance, consider a matrix transpose program.

\begin{SProg}
\qin\\
\qin transpose(Xs,[]) :-	nullrows(Xs).\\
\qin transpose(Xs,[Y|Ys]) :-	makerow(Xs,Y,Zs),	transpose(Zs,Ys).\\

\qin makerow([],[],[]).\\
\qin makerow([[X|Xs]|Ys],[X|Xs1],[Xs|Zs]):-	makerow(Ys,Xs1,Zs).\\

\qin nullrows([]).\\
\qin nullrows([[]|Ns]) :-	nullrows(Ns).\\
\qin\\
\end{SProg}

\noindent
As this program is intended to manipulate matrices of unknown type, we define 
the following types {\tt matrix}, {\tt row} and {\tt dynamic}, expressed as an FTA.  

\begin{SProg}
\qin\\
\qin [] -> matrix\\
\qin [row|matrix] -> matrix\\
\qin [] -> row\\
\qin [dynamic|row] -> row\\
\qin\\
\end{SProg}

\noindent
Our intention is to analyse the {\tt transpose} program in order to discover whether the program's
arguments have the expected types.  Note that the above FTA is not bottom-up deterministic
since there are two transitions with left hand side equal to $[]$.
Determinisation of this FTA together with the rules defining {\tt dynamic} for the program's signature,
yields
the following states and type rules (assuming that the signature is {\tt \{cons/2, []/0, 0/0, s/1\}}).

\begin{SProg}
\qin\\
\qin [] -> \{dynamic,matrix,row\}\\
\qin [\{dynamic,matrix,row\}|\{dynamic,matrix,row\}] -> \{dynamic,matrix,row\}\\
\qin [\{dynamic,row\}|\{dynamic,matrix,row\}] -> \{dynamic,matrix,row\}\\
\qin [\{dynamic\}|\{dynamic,matrix,row\}] -> \{dynamic,row\}\\
\qin [\_|\{dynamic,row\}] -> \{dynamic,row\}\\
\qin [\_|\{dynamic\}] -> \{dynamic\}\\
\qin 0 -> \{dynamic\}\\
\qin s(\{dynamic\}) -> \{dynamic\}\\
\qin\\
\end{SProg}

\noindent
Note that in the determinised automaton the states are sets of states in the original FTA.
These rules define an abstraction over the domain elements {\tt \{\{dynamic,matrix,row\},
\{dynamic,row\}, \{dynamic\}\}}, which represent the following sets of terms.
\begin{itemize}
\item
{\tt \{dynamic,matrix,row\}}:  the set of terms that are in all the types {\tt matrix}, {\tt row} and {\tt dynamic}.
{\tt \{dynamic,matrix,row\}} is
equivalent to {\tt matrix} since {\tt matrix} is a subset of both {\tt row} and {\tt dynamic}.
\item
{\tt \{dynamic,row\}}:  the set of terms that are in the types {\tt row} and {\tt dynamic} but not in {\tt matrix},
that is, it is the set of lists whose elements are not all lists.
\item
{\tt \{dynamic\}}:  the set of terms that are in {\tt dynamic} but not in the other types.
\end{itemize}
These three types are disjoint and complete; every term is in exactly one of these types.
We can compute the minimal model of a program based on this DFTA (for details of the process see \cite{Gallagher-Henriksen-ICLP04}).
In the model of this program over the domain {\tt \{\{dynamic,matrix,row\},
\{dynamic,row\}, \{dynamic\}\}} defined above, the {\tt transpose} predicate has the model

\begin{figure}
\begin{center}
\includegraphics[width=5.3 in]{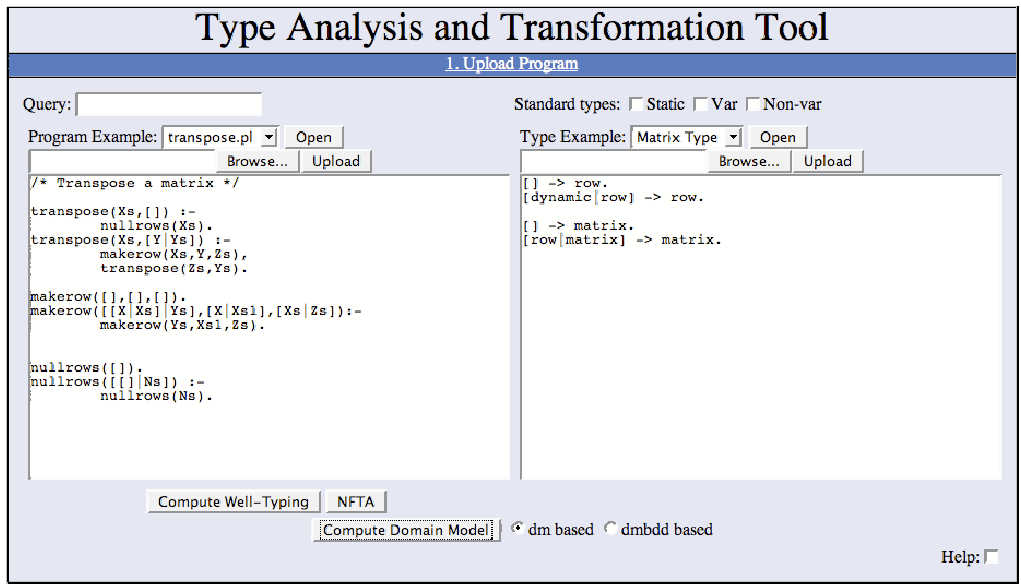}
\end{center}
\caption{The main tool interface showing a program and type ready for analysis}
\protect\label{fig1}
\end{figure}

\begin{figure}
\begin{center}
\includegraphics[width=5.3 in]{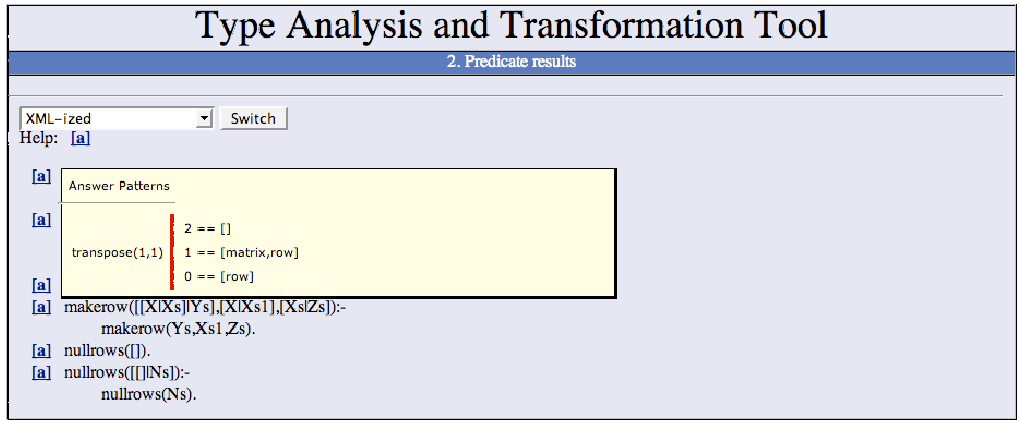}
\end{center}
\caption{Displaying the model of the {\tt transpose} predicate}
\protect\label{fig2}
\end{figure}

\begin{SProg}
{\tt transpose(\{dynamic,matrix,row\},\{dynamic,matrix,row\})}\\
\end{SProg}

\noindent
indicating that it can only
succeed with both arguments {\tt matrix}. 
Figure \ref{fig1} displays a screenshot of the tool, showing the {\em transpose} program, and the regular type
defining {\tt matrix} and {\tt row}.
Figure \ref{fig2} shows the results of computing the domain model, for the {\tt transpose}
predicate.  The model of the predicate appears when the mouse is moved over the symbol
to the left of the head of a clause for {\tt transpose}.  Note that for brevity the domain elements are numbered and 
a key is given alongside the predicate's model.  Also, the type {\tt dynamic} is omitted since it intersects
with every other type, hence the type {\tt []} mentioned in Figure \ref{fig2} is in fact the type {\tt \{dynamic\}}.
If any predicate has an empty model, then the heads of its clauses are highlighted in red.
\subsubsection{Modes as regular types.}
The type {\tt dynamic} depends on the signature of the program.  We call such types {\em contextual} types
\cite{GallagherPA05}.  Other contextual types are those defining the set of grounds terms (called {\tt static})
or non-variables terms (called {\tt nonvar}).  We assume that the signature contains some constant
{\tt \$VAR} that does not appear in any program or query. The rules for contextual types are generated
automatically by the system for the given program's signature, and the user does not in fact see these
rules at all. The rules defining {\tt static} are all those of the form
{\tt f(static,...,static) -> static} for each function {\tt f} in the signature apart from {\tt \$VAR}, while the rules defining {\tt nonvar}
are all rules of the form {\tt f(dynamic,...,dynamic) -> nonvar} for each function {\tt f} in the signature
apart from {\tt \$VAR}. The rules for {\tt dynamic} do include a rule {\tt \$VAR -> dynamic},
and thus the types {\tt static} and {\tt nonvar} are not identical to each other or to {\tt dynamic}.
A type {\tt var} can also be defined using the single type rule  {\tt \$VAR -> var}.

In the tool, the user can select one or more of the standard types {\tt static}, {\tt nonvar} and {\tt var} and 
add them to the types to be used for analysis.  (The type {\tt dynamic} is always included automatically, to
ensure that the types are complete).

The tool can thus be used for some classic mode analyses.  By selecting the type {\tt static} alone, the
domain model computed represents the {\sc pos} analysis \cite{Marriott-Sondergaard-LOPLAS93}.  Determinisation of the types
 {\tt static} and {\tt dynamic} results in two elements that represent the sets of ground and non-ground terms
respectively.  Similarly, selecting the type {\tt var} (or {\tt nonvar}) alone results in a freeness dependency
analysis, distinguishing variable and non-variable terms. Combinations of given 
types such as {\tt list} with {\tt static} or {\tt nonvar} allows the user to
distinguish say, ground and non-ground lists, or distinguish lists from other non-variable terms.

\subsection{Computing a Well-Typing}

\begin{figure}
\begin{center}
\includegraphics[width=5.3 in]{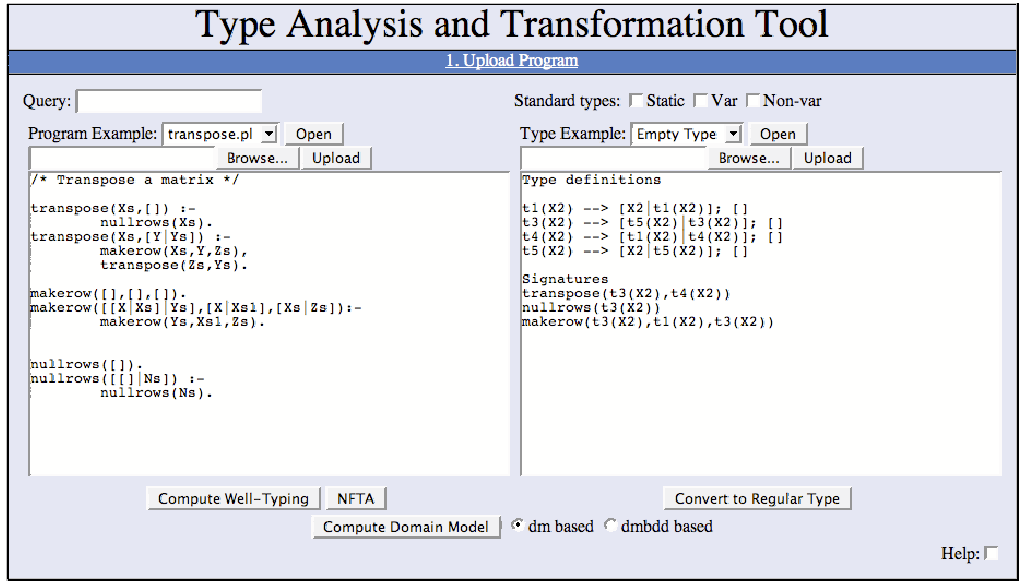}
\end{center}
\caption{Displaying a well-typing of the {\tt transpose} program}
\protect\label{fig3}
\end{figure}

The well-typing tool computes a set of type rules and signatures and is described in detail in \cite{BruynoogheGH05}.
The result is displayed in the type window on the right hand side of the screen (see Figure \ref{fig3}).
The types are parametric, and thus are more expressive than regular types.  Note that the types inferred
for the {\em transpose} program are in effect the types {\tt matrix} and {\tt row} defined earlier, except that
there is a type parameter representing the type of the elements of the rows.  
Recall that a well-typing does not necessarily represent a safe approximation of the success
set of the given program.  It only gives types that consistently indicate the way in which
the predicates are actually called in the program.  As we will discuss in Section \ref{sec:desctopresctypes}
the well-typing can be checked to see whether it is also a safe approximation.
Note that there is
some duplication in the types; for example {\tt t3} and {\tt t4} are renamings of each other,
as are {\tt t1} and {\tt t5}.   In future work we intend to identify and remove such identical types, using
the determinisation algorithm employed in the domain model tool to detect the redundancy.

\subsection{Computing a Non-Deterministic Regular Type Approximation}

The third analysis method
is the computation of a non-deterministic finite tree automaton
that over-approximates the least model of the given program.  This method is described in \cite{Gallagher-Puebla-02}.
The generated types tend to be complex and difficult to read.  The most interesting 
information is usually the emptiness of a type.  
As with the well-typing, the result is displayed in the type window, and can then be converted
to a regular type and used to build a domain model.  Conversion to regular type form in this
case is simple, as the inferred types are already regular type rules. We just need to remove
the rules defining the types of the predicates (which will be recomputed during the domain
model construction).

\section{Goal-Dependent Type Analysis}\label{qa}

\begin{figure}
\begin{center}
\includegraphics[width=5.3 in]{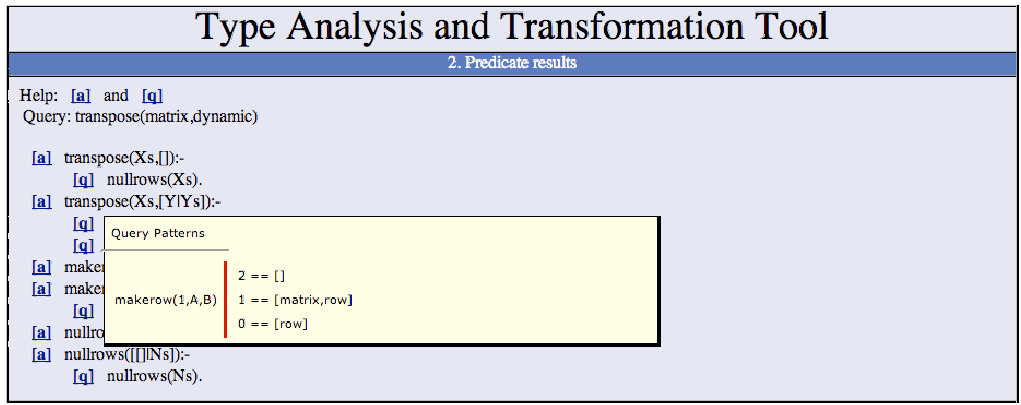}
\end{center}
\caption{Displaying query patterns}
\protect\label{fig4}
\end{figure}

Transformations to allow goal-dependent analysis using a goal-independent analysis tool are 
well known \cite{Debray-Ramakrishnan-94,Gallagher-deWaal-LOPSTR92}, and have their origin in the ``magic set" and ``Alexander"
methods from deductive databases\cite{Bancilhon-Maier-Sagiv-Ullman}.   The common feature of these transformations is the definition
of ``query predicates" corresponding to the program predicates. The variant we use in the tool 
constructs a separate query predicate for each body literal in the program.  A similar transformation
was described in \cite{Gallagher-deWaal-LOPSTR92}.  As the result of analysis of query-answer transformed programs can
be hard to read, the tool displays the models of the query predicates at the corresponding body
calls in the original program.  An example is shown in Figure \ref{fig4}.  When the mouse is moved
over the symbol to the left of each body call, the query patterns for that call are displayed, along with
a key to the determinised types, as before.

\section{From Descriptive to Prescriptive Types}\label{sec:desctopresctypes}

As mentioned earlier, a descriptive analysis such as the well-typing or the NFTA analysis constructs
types automatically.  Apart from providing documentation about the program predicates, these types 
can be used for prescriptive analysis where types have to be provided.
In \cite{BruynoogheGH05} the types were used to generate type-based norms for termination analysis.
In the present context, the inferred types can be input to the domain model tool.
The main reason for doing this is to generate more precise information about failure and dead
code in the program.  Consider the {\em naive reverse} program.  The well-typing analysis informs
us that both arguments of {\tt reverse} are lists, but we do not know whether this is a safe approximation.

A goal-independent NFTA analysis of the same program gives the information that the second argument
of {\tt reverse} is approximated by {\tt dynamic}.  By taking the result of either of these analyses and
using it as the regular type in the domain model tool, we can verify that the second argument of
{\tt reverse} is indeed approximated by the type {\tt list}.  (We would get the type presented
with some system-generated type name such as {\tt [] -> t1, [dynamic|t1] -> t1} which we would have to
recognise as {\tt list}.  Recognition of standard regular types is already performed by the
{\tt CiaoPP} analysis tool \cite{Hermenegildo-Bueno-Puebla-Lopez} and might be added to our tool in the future).
This allows us to detect as ill-typed any call to {\tt reverse} whose second argument is typed by
a non-list.  

As can be seen in Figure \ref{fig3}, the tool offers the possibility of using the inferred well-typings
or NFTA analysis results
to construct regular types.  In the case of well-typing rules this just involves replacing the parameters by {\tt dynamic}, 
which involves a loss of precision.  The NFTA types can be used directly with minor syntactic
changes. The regular types
can then be used to build a domain model, as in Section \ref{dm}.

NFTA types tend to be large and complex with much redundancy in the form of multiple occurrences
of the same type with different names.  So far, the most likely uses of NFTA analysis is to
find useless clauses, dead-code, failing calls and such like.
Determinisation and computation of a domain model can enhance the ability to search for
such anomalies.  An example is provided by the {\tt tokenring.pl} program on the sample program
menu.  An NFTA analysis
yields types, but every predicate has a non-empty type.  If the derived types are converted to
a regular type and a domain model is computed, then the fact that {\tt unsafe/1} has an 
empty type is detected.  The BDD-based domain-model tool is required to get this result;  the
NFTA type rules are complex and the determinised automaton has 111 distinct states.

\section{Features of the Implementation}\label{impl}

The Type Analysis Toolkit consists of a back-end and a front-end.
The back-end consists of the analysis programs themselves, and the front-end 
is a user friendly web-interface for those tools. There are obvious 
advantages of using a web-based interface; one of these is of course 
that anyone wishing to try the tool for himself can do so without 
having to install software on his own machine and dealing with possible 
incompatibilities in operating systems and so on. It is also intuitively 
easier to use a programming tool with a graphical interface, than having 
to e.g. read and decipher command line options.

\subsection{Back-end}
There are currently three tools in the back-end: the domain-model analyser, the
polymorphic type analyser and the NFTA analysis tool - all written in Prolog. 
The particular Prolog
system we are using is the Ciao Prolog Development System\footnote{{\tt http://clip.dia.fi.upm.es/Software/Ciao/}}.

Initially the back-end tools were developed independently and were intended to be
executed within a Prolog environment. They have been modified to allow
them to be compiled into two separate command line tools. It is also possible
to use the Ciao-shell environment for running the back-end tools from the command line. 
This method does not require
compilation of the tools, but it comes with a performace penalty. Table \ref{tab:analysisperformance}
shows the analysis time for a few select programs; the append program, Leuschel \& Massart's model checker for CTL formulas \cite{Leuschel-Massart-LOPSTR99}, and the token ring program mentioned earlier in Section \ref{sec:desctopresctypes}. The table compares a compiled version of the Domain Model (DM) program to a version of DM executed in the Ciao-shell environment. For larger programs the penalty of using Ciao-shell may be insignificant, but for smaller programs the compiled version is significantly faster.

\begin{table}
\begin{center}
\begin{tabular}{|l|c|c|c|}
\hline
Program & No. Clauses & Compiled DM & Ciao-shell DM \\
\hline 
append.pl    & 2 & 0.4 & 3.7 \\
tokenring.pl & 20 & 0.9 & 3.5 \\
ctl.pl		& 27 & 3.7 & 7.3 \\
\hline
\end{tabular}
\vspace{0.2cm}
\caption{This shows the analysis time in seconds for a few programs containing from 2 to 27 clauses}\label{tab:analysisperformance}
\end{center}
\end{table}

The tools were also modified to read their input and write output to files to simplify the development
of the web interface. The output from the polymorphic type analyser is plain text. This output will
be read by the domain model tool. The output from the domain model tool will
be parsed by the front-end, so we decided to format this
output in XML. The result from the domain model tool is a term containing the
analysed program with clause heads and clause bodies annotated with the 
query and answer patterns. This term is
not a syntactically correct Prolog program, and we have therefor written our own 
Prolog module to handle the output of terms in an XML structure. Tools for
outputting and analysing Prolog code in XML exists \cite{DBLP:conf/lpe/SeipelHH03},
but at the moment we only need to separate clauses and calls from annotations when
presenting the analysis results in HTML.

\subsection{Front-end}
The invoking of the back-end tools and the presentation of the analysis
results are handled by the front-end. The front-end has two main parts;
a page with a form to be filled in by the user, where the user uploads, types in or 
selects one of the supplied example programs and types, 
and a presentation page where the analysis results are displayed.
The front-end is built on the Apache webserver, the PHP scripting language and
libxml and libxslt.
The design of the pages is inspired by other online analysers for
logic programming developed in the context of the Framework 5 ASAP project\footnote{
{\tt http://www.clip.dia.fi.upm.es/ASAP/}\\
{\tt http://www.stups.uni-duesseldorf.de/\~{}pe/weblogen/} \\
{\tt http://www.stups.uni-duesseldorf.de/\~{}asap/asap-online-demo/}
}.

The PHP language allows us to create dynamic pages in which, for example, buttons can be
removed from the input-form if they do not apply to the current input given. It also
enables us to execute the command line analysis tools and run the output through
XSLT. XSLT provides a convenient way of automatically transforming the XML output into HTML
that can be displayed in a browser. Depending on whether the analysis is performed with
or without a query, the appropriate XSL style sheet is applied to the analyser output.

\subsection{User's guide}
Using the analysis tools involves the following steps for the user:
\begin{enumerate}
\item Supplying a program for analysis
\item Inferring or supplying a set of types for analysis
\item If applicable, supply a typed goal for a goal dependent analysis
\item Selecting back-end tool for the analysis
\end{enumerate}

A few example programs and types are available for a quick demonstration of the tool. Select the program {\em tokenring.pl} in the dropdown menu to the left and the type {\em Ring Types} in the dropdown menu to the right. In the query-field type {\em unsafe(dynamic)}. Make sure the dm-based back-end is selected next to the {\em Compute Domain Model} button. Then click this button. Notice that the predicate {\em unsafe} has no answers, meaning no unsafe state can occur in the token ring.

\subsubsection{Selection of Program and Analysis Method}
An example of the input page is shown in Figure \ref{fig1}. On the left side of the
screen the user can either paste in a program to be analysed, select one of the 
example programs or upload a local file containing a program.
On the right side of the screen, the user can supply the type, again either
from a local file, by selecting one of the provided type definitions or by typing 
directly into the panel.
At the bottom of the screen are the options to either run the Polymorphic Type Analyser, 
the Domain Model Analyser or the NFTA analysis on the given program. In the case of
the Domain Model, the Prolog implementation or the BDD-based tool {\tt bddbddb}
can be selected.
If the Polymorphic Type Analyser or the NFTA analysis is used, the result is shown to the right. A new option
to convert to regular type will appear.

\subsubsection{Display of Analysis Results}
An example of the output page is shown in Figure \ref{fig2}. The analysed program is
shown annotated with answer patterns and if a query was supplied to the analysis, also
a query pattern for the calls in the body of the program clauses.

Placing the mouse over either of the annotations will show a small window with the
actual patterns.

Should a clause have an empty answer pattern it will be coloured red to indicate
that it is dead code. If a query was given to the analyser, the code is considered
dead with respect to that particular query pattern.

Calls in the body of a query having an empty call pattern are similarly highlighted
in red. These are calls that are redundant; they and the calls to their right in
the clause can safely be ``sliced" from the program, since they are not invoked in the
computation of the given query.

\paragraph{}
The Analysis Toolkit (called {\sc Tattoo} -- Type analysis and transformation tool -- is available online - to try it out, visit the URL 
{\tt http://wagner.ruc.dk/Tattoo/}.

\section{Future Development}\label{future}

The current tool provides a platform for experimentation.  The front end and the back end are cleanly separated
and so further tools could be added to the back end with minimal modification of the front end.  The use of XML also
allows us to experiment with different ways of displaying the results to the user.

Analysis of large programs using complex types is likely to require powerful tools such as the BDD-based analysis tool described
in \cite{Gallagher-Henriksen-Banda-2005,bddbddb}.  An option in the interface to use this tool rather than the Prolog implementation
has been implemented, but only goal-independent analysis can be carried out with this tool currently. 

So far our effort has been focussed on developing the tool and considering the best way to present the
results to the user and allow the results of one tool to be input to another.  The interface will continue to
undergo development as more experience is gained.  Evaluation of the tool is being carried out and
we hope that its
availability on the web will enable feedback to be obtained from a wider range of users.

\subsection*{Acknowledgements}

We thank the ASAP project team ({\tt http://www.clip.dia.fi.upm.es/Projects/ASAP/}) for collaboration during the
development of the tools and the online ASAP tool interface which incorporates some of the type analysis tools
described here.

\bibliographystyle{splncs}
\bibliography{refs}

\end{document}